# A Sociotechnical Readiness Level Framework for the Development of Advanced Nuclear Technologies

Aditi Verma[a] and Todd Allen[a]

[a] *Nuclear Engineering and Radiological Sciences, University of Michigan, 2355 Bonisteel Boulevard Ann Arbor, MI 48109, USA.*

[*]aditive@umich.edu

# A Sociotechnical Readiness Level Framework for the Development of Advanced Nuclear Technologies

The Technology Readiness Level (TRL) scale was initially developed by NASA in the 1970s and is now widely used in space, nuclear, and other complex technology sectors in the US and beyond. The TRL scale is particularly useful for determining where extrapolation of untested sub-systems or features could produce technical risk, cause expensive redesigns, or act as a roadblock to technology development. In this paper, we propose the development of a sociotechnical readiness level (SRL), premised on the understanding that the successful development and eventual use of a technology requires achieving not only full technological readiness but also anticipating, prioritizing, and addressing societal concerns that may arise during the course of development of a technology. Failures to anticipate and address societal factors in the early stages of technology development have led to high-profile delays and, in some cases, ultimate failures of nuclear technology projects. The sociotechnical readiness scale, which conceptually draws on the design research and science and technology studies scholarship, centers on principles of equity and environmental justice in technology design, and emphasizes the need for social engagement during the process of technology development. Nowhere is such an approach to technology development more vital or needed than for the long-term management of spent nuclear fuel.

## INTRODUCTION

Designers, planners, and policymakers in the US and globally are considering and developing energy technologies for the rapid decarbonization of energy systems before mid-century. Within the context of energy, climate, and policy discussions, nuclear energy is increasingly being regarded as an important, potentially critical, contributor to our future low-carbon energy systems. However, previous generations of nuclear technologies have created a range of equity and environmental justice challenges. These challenges – which span the nuclear fuel cycle – include the displacement of indigenous communities to site legacy nuclear facilities, failures to engage host communities as part of the development, siting, and remediation processes for such facilities, contamination of land and water resources in the event of technological failures (not limited to nuclear accidents), the absence of a long-term plan for the stewardship of nuclear waste leading to its indefinite storage at nuclear plant sites, as well as a failure on the part of the nuclear sector and nuclear professionals to inform communities about the benefits and harms of these technologies and securing informed consent prior to the development of these facilities. These inequities and environmental injustices have both eroded trust between the nuclear sector and host communities and impacted the potential scale and pace at which nuclear energy technologies may now be adopted for decarbonization purposes.

Many of these equity and environmental justice challenges can be addressed by transforming how we approach technology development processes. Specifically, by incorporating community input and perspectives in technology design and siting, community concerns and preferences can be incorporated early and throughout the technology development stages. This can increase the likelihood that the technology and facility developed and sited is viewed by the host community as desirable, representative of community values and input, and developed on the basis of consent that was given in a way that is not coercively extracted. Enfolding these values in technology development requires a fundamental reimagination of energy technology development processes.

As a step in the direction of enabling inclusive and equitable energy technology development – we propose a reformulation of the Technology Readiness Level (TRL) frame. The TRL scale is used to assess the maturity and completeness of a complex technology or system. The TRL scale was initially developed by NASA in the 1970s and is now widely used in space, nuclear, and other complex technology sectors in the US and beyond. The TRL scale is particularly useful for determining where extrapolation of untested sub-systems or features could produce technical risk, cause expensive redesigns, or act as a roadblock to technology development. In this paper, we propose the development of a sociotechnical readiness level (SRL), premised on the understanding that the successful development and eventual use of a technology requires achieving not only full technological readiness but also anticipating, prioritizing, and addressing societal concerns that may arise *during* the course of development of a technology. Failures to anticipate and address societal factors in the early stages of technology development have led to high-profile delays and, in some cases, ultimate failures of nuclear technology projects – both waste and energy. The sociotechnical readiness scale, which conceptually draws on the design research and science and technology studies scholarship, centers principles of equity and environmental justice in technology design, and emphasizes the need for social engagement *during* the process of technology development. Nowhere is such an approach to technology development more vital or needed than for the long term management of spent nuclear fuel.

Prior attempts to expand the technical readiness level framework to include societal elements such as market, regulatory, and public readiness have treated these as the ends rather than a means to equitably develop technologies and have demarcated the pursuit of these aspects of readiness as parallel processes separated from technology development. In this paper, we conceptually develop the sociotechnical readiness level, which, across each of its nine levels emphasizes procedural, distributive, recognition, restorative, and epistemic forms of justice and equity.

We expect that the sociotechnical readiness framework can be used to inform the development and assessment of advanced nuclear reactor technologies at various stages of the advanced fuel cycle and, more broadly, the development and assessment of complex clean energy projects. We aim to explore these applications of the SRL in our future work and invite researchers and technology developers to do so as well.

**THE TECHNOLOGY READINESS LEVEL FRAMEWORK**

Technology Readiness Levels (TRL) are a measurement system developed to assess the maturity level of a particular technology and manage technical risk throughout the product development process of new and complex technologies. It was originally developed in the 1970s at NASA. Early applications of the TRL framework include readiness assessments of planet atmospheric entry probes [1], of atmospheric revitalization technologies for manned spacecraft [2], and a space station [3]

The TRL framework was later adapted and broadly deployed across NASA, the U.S government (Government Accountability Office, Department of Defense, Department of Energy), and internationally as a standard means to measure technology maturity and its advancement over the course of product development. In both U.S. Departments of Energy and Defense, TRLs are tied with critical decisions and the funding life cycle of new technologies to ensure that projects do not overlook technology immaturity in a manner that can lead to costly schedule overruns or project failures.

At the lower end of the scale (TRL=1-2), scientific research is just beginning and only analytical calculations exist to support the feasibility of the concept. Once experimental data related to the performance of a system is collected and proves the concept, the TRL reaches a level of 3. The TRL advances from 4-6 as experiments are done with increasing similarity in scale to the final application. At TRL= 7-9, the system prototype is operated over a broader set of environmental and operational conditions that eventually match the actual system.

The technology readiness level approach breaks down a system into its components and the TRL of the system is determined by the lowest TRL of the components. Thus, the TRL scale can fail to properly identify important interactions and challenges associated with integrating two or more technologies that may be mature in their own contexts but may not be mature when combined in novel configurations. Alternative approaches such as the "system readiness assessments" have been proposed to address this limitation [4].

Recent attempts to expand the concept of technical readiness have led to the development of the regulatory and market readiness levels [5], and more recently, the balanced readiness level assessment scale [6] which, along with technical readiness, emphasizes market, regulatory, public acceptance, and organizational readiness, thus separating these aspects of readiness from technological readiness and identifying these categories of readiness as meriting separate but parallel consideration.

The regulatory and market readiness level framework appends regulatory and market considerations to those of technological readiness. While the technical readiness framework, as described above, inquires about each stage of technological development starting with the observation of basic scientific principles through to the qualification and proof of concept of a system, the regulatory and market readiness levels – each of which are divided into five levels corresponding to various combined levels of technological readiness – inquire about the status of regulatory approvals and market potential of the technology. The initial stage of the regulatory readiness level, which is concerned with access to the regulatory process, corresponds with the first two levels of technological readiness. The next regulatory readiness level is about security of political capital, followed by policy effectiveness, safety (phrased as "do no harm"), culminating with political acceptability. Similarly, the market readiness levels begin with access to a market base, followed by security of financial capital, manufacturing capacity, an assessment of the profitability of the technology in question, and culminating with consumer utility. Both the regulatory and market readiness levels, as they are conceptualized, center the technology developers and their imperative to successfully and rapidly develop a technology, rather than the needs of the users of the technology or those who may be impacted by it. For example, the market readiness levels culminate with consumer utility, rather than starting with an assessment of consumer utility as a precondition for considering whether a technology ought to be developed in the first place, and if it should, what form and functions the artifact or system under development ought to take. Similarly, the regulatory readiness levels appear to be concerned with building a regulatory framework and environment that offers certainty to the technology developer across each stage of technological development, rather than offer consumers and those potentially impacted by the technology being developed access to both regulation as well as the technology development process itself. These considerations of consumer (or taken at larger scale, community and public) utility and safety, if only included as after-the-fact or late-stage considerations, though possibly resulting in the rapid development of a new technology, are ultimately unlikely to lead to the development of a technology which is equitably designed as well as one that equitably distributes benefits (and any potential harms) resulting from it.

The balanced readiness level assessment, adds two further elements of readiness to the previous triad – these are the acceptance readiness levels and the organizational readiness levels. Here, the acceptance readiness level is concerned with the legitimization of a new technology or its "social acceptance". The acceptance readiness levels assume that a technology, when it is first conceived, will inevitably be viewed as illegitimate and not be accepted widely or at all in society writ large. According to this framework, the acceptance readiness of a technology ought to increase along with its technological readiness until, in the later stages of acceptance readiness, the "use and production of the technology is generally accepted and not questioned at all"[6]. Under this framework, increasing acceptance is secured not as a result of material changes to the technology but rather through a campaign of public education or a happy byproduct of increasing regulatory readiness. Similarly, the organization readiness level is concerned with the "domestication" of a technology and an assessment of the compatibility of the new technology with existing ones and therefore their potential for integration – as appropriate. In the initial stages of organizational readiness, the technology is clearly incompatible with existing technologies and organizational practices. However, this compatibility, in an ideal case, is expected to increase until the new technology is integrated both with existing technologies as well as work practices and processes. As with the other non-technological forms of readiness, the organizational readiness levels are assumed to be separate objectives pursued parallel to the design and development of a technology but having no impact on the technological design and development itself. In other words, the framework appears to assume that it is the adapting organization that must conform to the new technology, rather than the technology being designed to be fit for existing users and circumstances.

Both frameworks described above have a number of limitations which we have alluded to above. These include:
(1) Conceptualizing technological design and development as a strictly linear process and similarly conceptualizing other forms of readiness (market, regulatory, acceptance, organizational etc) as also proceeding in a linear fashion
(2) Pursuing technological readiness separately from other aspects of readiness, with other aspects of readiness – such as market, regulatory, and organizational readiness – having little to no bearing on how the technology in question may be designed;

(3) Viewing users, communities, and publics and their concerns as peripheral to the technological development process and having no bearing on or access to the design of the technology

(4) Excluding consideration of histories of development of similar technologies, including harms caused by such technologies, and excluding a consideration of how past histories and harms ought to inform future development, including restorative work, and trust building, wherever appropriate

(5) Excluding perspectives (even if just speculative) of future users, communities, and publics from considerations of readiness. Such considerations are especially important for long-lived technologies and systems whose lifetimes and remains are expected to outlive any single generation, including certainly the generation that develops the technologies.

We view societal aspects of readiness as being fundamentally inseparable from technological readiness and instead propose the sociotechnical readiness level (SRL) framework for considering these aspects of readiness within and throughout the process of technological development, as opposed to separate processes as done by the existing frameworks described above. The SRL framework is conceptually motivated by design research and science and technology studies scholarship which is described next.

**CONCEPTUAL BACKGROUND**

Most engineering disciplines, including nuclear engineering, view design as a central skill and output of a field or discipline but do little to theorize the structure of the design process or examine what factors and constraints – beyond physical constraints – shape design choices[7–11]. Elsewhere, in the design research [12–14], history of technology [15–17], and science technology and society fields [18–20], there has been a long history of studying design and making both the designed artifact as well as its process of development objects of study.

Recent scholarship in nuclear engineering [21], as well as long standing work in the field of design research writ large finds that the designers' expertise, background and identity, the organizational site of the design work as well as several other factors such as the form of funding available to designers, tools used during each stage of design all non-trivially shape the design choices that are made [22–24]. This work also affirms what most practitioner designers know to be experientially true – that design is not a strictly linear-analytic process starting with problem identification and culminating with solution and that the decisions made in the early conceptual stages of design often irreversibly influence the final form that an artifact or system takes. This research dispels the myth of linear, staged technological development and implores designers to reflect on the complexity and nuance of their own practice, including reflecting on openings in their work for the input, and in some cases, direct participation of users and communities

[25,26]. These approaches and methodologies for enfolding community and user input in the design process are variously referred to as human-centered design [27], user-centered design [28], resource-constrained design [29], designing for the bottom of the pyramid [30], frugal design [31], values in design [32] and design justice [26].

Technologies can span many nested scales with the smallest level of scale being a technology concept, which in turn may be instantiated in a concrete artifact, which forms part of a larger system, which in turn may be a sub-system in a complex sociotechnical system [33]. It is possible, over the course of technology design, development, and distribution for inequities to become embedded – intentionally or unintentionally through the decisions made by the technology developers and designers at any (or even all) of these levels of scale. These inequities, if left unexamined, can be perpetuated whole cloth to future generations of that technology or system as well as others that may emulate it. Preventing such an entrenchment and perpetuation of inequities requires, according to the design justice framework [26], that technology designers and developers – at each level of technological development (in our case, technological readiness) – ask a series of questions about who is privileged to design technologies, who technologies are designed for and with, what values become embedded (intentionally or otherwise) in technological designs, how the sites of technological design work are impacted, and how we rationalize and stabilize technologies once they have been created. Asking these questions in the context of energy technology development can enable a pursuit of procedural, distributive, restorative, recognition, and epistemic forms of energy and environmental justice [34,35].

Our reformulation of the technological readiness framework as the sociotechnical readiness framework marks a move away from viewing technology design and development as a linear process. It accounts for the multiple scales that technologies may span, and considers inequities, as perceived by present and future users and communities, that may become embedded across those scales. Finally, the sociotechnical readiness framework does not view new technologies as ahistorical but rather situates them in their technological lineage and inquires what harms and burdens may have been created as a result of the development and use of previous technologies. According to this framework, the various forms of readiness – market, organizational, regulatory, and acceptance, are not pursued in parallel with technological development but are instead, where they may result, the direct consequences of engaging publics and communities throughout the technological development process.

**THE SOCIOTECHNICAL READINESS LEVEL FRAMEWORK**

The main contribution of this paper is in the form of the sociotechnical readiness level framework which is laid out in Table 1 below. The sociotechnical readiness level framework as described below was developed by: (1) responding to the equity-centering critiques – as described above – of the traditional technological readiness level framework as well as the more recently developed regulatory, market, and balanced readiness level frameworks; (2) conceptually motivated by insights from design research, history of technology, and science and technology studies, and (3) augmenting the traditional technology readiness framework that is extensively used by the DOE and other federal agencies. In particular, while developing the descriptions of each level of SRL, we drew on the TRL descriptions as presented in Technology Readiness Assessment Guide [36].

The SRL framework is based on three normative principles:
(1) An inclusive technological design and development process that accounts for local expertise and knowledge during every stage of technological development and recognizes these local forms of knowledge as held by user and community groups as forms of expertise as vital to technological design and technological readiness
(2) An assessment of the actual and potential social and environmental harms caused and inequities created by prior versions of a similar technology or prior technologies performing a similar function, and a retroactive and continuing repair of those harms and inequities
(3) An anticipation of the futures resulting from the development and use of a technology – including the social, environmental, political, and cultural implications – and imagining those futures in inclusive ways

Figure 1 visually depicts the sociotechnical readiness level framework and Table 1 below presents nine levels of the sociotechnical readiness framework with descriptions of each level. The SRL framework seeks to integrate the technical and social aspects of readiness at each stage of *sociotechnical development* starting from the inception of a new technology.

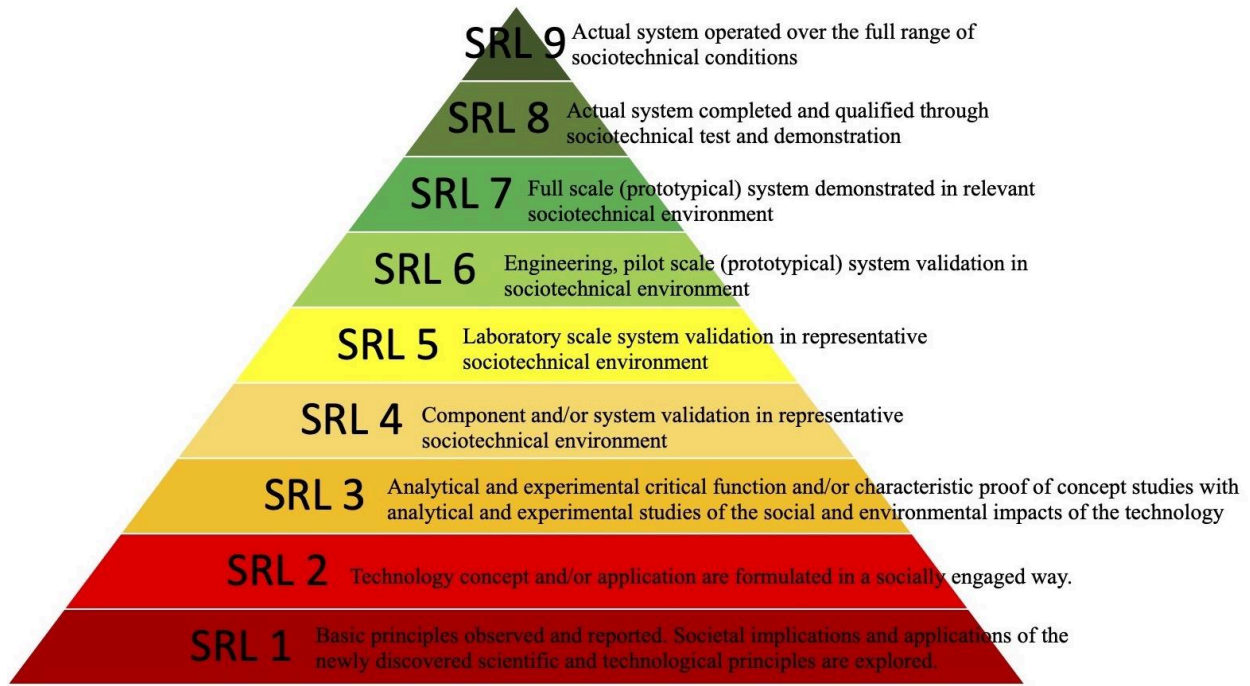

Figure 1. The sociotechnical readiness level framework

Table 1. The sociotechnical readiness levels and descriptions. The SRL descriptions were developed by building upon the TRL descriptions presented in the DOE Technology Readiness Assessment Guide [36].

| SRL levels | SRL definition | SRL Description |
|---|---|---|
| SRL 1 | Basic principles observed and reported. Societal implications and applications of the newly discovered scientific and technological principles are explored. | This is the earliest level of sociotechnical readiness during which applied R&D begins. This includes scoping sociotechnical studies to explore what real-world problems or challenges that the technology may help solve – as well as how these problems and solutions may vary across different demographics and communities – as well as analytical studies of the basic properties of a technology. Early-stage experimental work consisting of observations of the physical world may also be carried out. Supporting information includes published research and references that identify potential societal uses and applications of the technology (through direct |
| | | engagement with potential users and communities) and identify the scientific and technical principles underlying the technology. |
| SRL 2 | Technology concept and/or application are formulated in a socially engaged way. | Practical applications of the technology based on user, community, and societal needs (as identified in SRL1 exploratory studies) are invented. Though applications may still be speculative, the majority of potential applications are backed by empirical data collected from potential actors and stakeholders in SRL1 exploratory studies. Supporting information includes publications, references, and studies that support the sociotechnical validity of the concept.

The sociotechnical validity of the concept is supported by establishing the technical validity of the concept through experimental and analytical studies with an emphasis on understanding the science better. The social validity of the concept is supported by empirical qualitative and quantitative research done in collaboration with users and communities. The purpose of the user and community-focused studies is to (1) Attempt to exhaustively identify potential applications of the technology in collaborations with users and communities, (2)Assess how the benefits and harms resulting from the different potential applications will be distributed socially (across a wide range of markers of identity – race, socioeconomic status, gender, disability etc) and environmentally across all time scales, and (3) Assess whether other technologies used for similar application in the past have created social or environmental harms and whether those harms have been repaired. If those harms have not yet been repaired, plans to repair those harms are formulated as part of the technology development process.

This transition from SRL 1 to SRL 2 marks a shift from pure to applied research and establishes the sociotechnical validity of the proposed technology. |
| SRL 3 | Analytical and experimental | Active research and development is initiated in this phase. This includes socio-analytical design studies to develop more detailed technological |
| | critical function and/or characteristic proof of concept studies with analytical and experimental studies of the social and environmental impacts of the technology | designs. These socio-analytical design studies involve the direct participation of potential users and host communities. While these users and host communities may not be technical experts per se, they can offer invaluable input about the social and environmental implications of design choices as well as stimulate the exploration of new design choices that serve societal needs – thus potentially enabling new and unforeseen technological applications and markets. Facilitation of this designer-user/community dialog requires that the technology developers develop an accessible language for describing design choices and their socio-environmental implications. Crucially, this phase also involves the laboratory scale studies to physically validate the design choices developed through the socio-analytical design studies. Laboratory scale studies validate the separate elements of the system and not yet the sociotechnical system as a whole. Supporting information includes the results of the socio-analytical design studies as well as the laboratory studies published in the form of reports, articles, or other forms as appropriate. |
| SRL 4 | Component and/or system validation in representative sociotechnical environment | The technical components of the system are integrated to establish that they will work together and the sociotechnical interfaces of the system are identified and sociotechnical integration is also performed. While some complex systems may only have technical parts and may be truly autonomous, most complex systems are sociotechnical systems. For example, a nuclear reactor is a sociotechnical system because human operators, maintenance staff, and security personnel are key to the safe, secure, and efficient operation of the system. As yet this is a low-fidelity compared with the eventual system that will be developed. SRL 4 marks the first step towards determining whether the individual technical as well as social components will work together as a system. |

| SRL 5 | Laboratory scale system validation in representative sociotechnical environment | The sociotechnical components of a system are integrated so that the system configuration is nearly identical to the final application. Examples include testing a high-fidelity, laboratory scale sociotechnical system in a simulated environment – including with potential users, with a range of simulants, and actual wastes. The results of these studies are shared with the user and community groups who have participated in studies carried out in prior SRL levels and their reactions and inputs to the laboratory scale studies are solicited. Supporting information includes results from the laboratory scale testing, analysis of the differences between the laboratory and eventual/completed system, a synthesis of the inputs from the user and community groups, and an analysis of what the experimental results as well as user and community inputs mean for the design of the final system. |
|---|---|---|
| SRL 6 | Engineering, pilot scale (prototypical) system validation in sociotechnical environment | Engineering-scale models or prototypes are tested in a relevant sociotechnical environment with direct participation or observation of user and community groups involved in prior SRL phases. Examples include testing the engineering scale prototypical system with a range of different simulants, a range of different users, and in the context of a range of different potential host community environments which may be virtual or real. Community scale testing at this stage may be facilitated by the use of immersive experiences such as virtual or augmented reality technologies or other formats identified collaboratively with the participating communities.<br><br>Supporting information includes the results of experimental and sociotechnical studies published in the form of reports and articles. The major difference between SRL 5 and 6 is the shift from laboratory scale to engineering scale. The sociotechnical system developed and demonstrated during this phase must be capable of performing all functions and meet the applications collaboratively identified with user and community groups. |
| SRL | Full scale | This phase requires the demonstration of an actual system prototype in |

| 7 | (prototypical) system demonstrated in relevant sociotechnical environment | one or more relevant sociotechnical environments. Examples include testing a full scale prototype by building it in a potential host community and with the hands-on involvement of users. Additional users and host communities who have participated in prior levels of SRL work but who are not directly involved in this stage of work, may participate as observers and provide input. Social and environmental impacts of the prototype on the demonstration site are documented as exhaustively as possible with continued and periodic documentation of these impacts after the prototyping work is completed. Supporting information includes findings from the prototype construction, demonstration, and dismantlement, input from the users and community groups involved in this process directly or as observers, as well as a synthesis of what all of these findings imply for the final system design. |
|---|---|---|
| SRL 8 | Actual system completed and qualified through sociotechnical tests and demonstration | The sociotechnical system has been proven to work in its full-scale form. However, this may not yet be the final form for the system as it may continue to be improved as it is operated and as future versions of the system are designed and built. For example, as the system is used, new information may be revealed about its internal or external sociotechnical interfaces that may merit a reassessment of integration. Further, just as preventative maintenance is carried out on technical systems, the system developers ought to continue to monitor the sociotechnical interfaces and update the external community-system interfaces in consultation with the community.<br>Examples of work carried out during this stage include developmental testing and sociotechnical evaluation – including plans to manage both front-end extractive activities and waste management in a socially and environmentally responsible way. |
| SRL 9 | Actual system operated over the full range of sociotechnical conditions | The technology is in a form in which it can be built and operated while having accountability - to the extent possible – for the social and technical improvements identified in earlier stages of development. As previously noted, there is not necessarily an absolute final form of the technology, nor should the development of a final form be the end goal. The process |

| | | of technology design and development is by no means linear and novel ideas for improvement may be identified throughout the development of a technology. Bearing this in mind, any such ideas for improvement – whether their provenance be from the design team or from the user and community groups – have been codified and made available to the designers working on the next iteration of the technology such that the current iteration is not reified. |
|---|---|---|

**DISCUSSION**

One possible critique of this approach that may be offered by the engineer-designer is that users and communities who do not have scientific knowledge or expertise may not be able to provide input to the design process. While it is true that members of a community may not directly be able to call for certain reactor design choices or critique others – having to do, for example, with the type of fuel used, the operating temperatures and pressures, or the balance of plant – they are likely to have significant input if presented with the socio-economic and environmental implications of those choices. For example, though micro and small modular reactor designers have increasingly been placing an emphasis on autonomous designs and drop-in concepts – this may in fact be undesirable to host communities who may wish to benefit from high skilled and high-paying jobs in reactor operations and construction (as has traditionally been the case in communities hosting nuclear energy facilities). It therefore falls to the designers of the technology to engage with potential users and communities using a vocabulary and language of design choice implications rather than jargon-laden descriptions of the artifacts and systems being designed.

While user and community-centric design approaches have not yet been adopted for the design of complex systems, there has been an increasing movement towards the adoption of such approaches in the design of automobiles, web applications, software systems, personal robots, and breast pumps (to name some). Application of inclusive and participatory design approaches in these contexts is made possible by the existence of a well-defined user or set of users, and a clear relationship between user desirability and product or system success. Complex sociotechnical systems however do not generally have well-defined users. Instead these systems, at least historically, have been designed by business and organizations for other organizations. Even where users are present, they are regarded as being part of the system (nuclear plant operators) rather than individuals with agency comparable to a customer-user. Despite not having users, complex sociotechnical systems have stakeholders – a diverse set of actors, who are impacted by the design, development, and use of the system – and the impacts on these stakeholders may be heterogeneous depending on their socioeconomic status, race, gender, ethnicity, disability and other defining features of identity. Further, while these stakeholders may not necessarily be implicated in the day-to-day operations of a system, they are likely to be significantly impacted in the case of a large-scale system-wide failure – a nuclear plant accident, the failure of an electric grid, the collapse of a bridge, for example. As noted above, these stakeholders, though typically not consulted in the process of technology development, likely have context-specific invaluable input to provide to the technology developers.

Technological design and invention often result from serendipitous and playful exploration of the potential applications of scientific principles. Through the development of the SRL framework we, by no means, seek to discourage this serendipitous exploration and discovery, but rather seek to encourage technology developers to invite potential users and those who may be impacted by a technology – positively and negatively – into the very process of exploration and play. Doing so is likely to not only ultimately lead to technologies that are intersectionally and widely useful but also open up new parts of the design space the science-focussed designers may not have thought to explore. Bringing users and communities into the process of technology design calls for the development of new modes and processes of design and engagement that make design a more accessible and inclusive enterprise and open up the definition of who has design expertise and who is regarded as a designer. Ultimately, these user and community-centered modes of technology development may in fact slow down the early stages of the development of a new technology or system but are likely, in the long run, to (1) prevent the adoption of socially and environmentally undesirable design choices that are pursued into late stages of technological development and thus are difficult to recover from, leading potentially to project failure, (2) prevent the expensive pursuits of unwanted technologies and, (3) potentially avoid the delays, cost overruns, and oppositions at initial deployment and also throughout the life of a system in which inequitably designed technologies – including nuclear technologies – frequently become mired. Through the SRL we are proposing user and community input as being an early, logical, ethical, and mutually beneficial part of the design process.

**APPLICATIONS OF THE SOCIOTECHNICAL READINESS LEVEL FRAMEWORK**

This paper has predominantly focused on the motivation for, development of, and discussion of the SRL framework. We now briefly discuss two potential applications.

**Advanced reactor design**

Nuclear reactors, since their inception and initial development have been scaled up from under 100 MWe to large, increasingly complex systems of 1000 MWe and greater and have traditionally been intended as grid-scale sources of energy. However, several advanced reactor developers have been pursuing an opposite trend – scaling down these systems in size and complexity – to small modular, micro, and even nano reactors intended to have both electric and non-electric applications, as well as the potential to be used off-the-grid including in remote communities. This fundamental shift in the design paradigm also calls for a shift in technology design and development approaches. Sociotechnical systems such as nuclear reactors, though they may have only a handful of 'users' in the form of operators – have numerous and often diffused stakeholders who are impacted by the development, use, and dismantlement of such systems as a result of living around the sites of development. Including the input and perspectives of these stakeholders is especially important for the developers of advanced reactors who, in many cases, aspire to embed these reactors at the community level. The SRL provides a framework for structuring and assessing this community engagement and calls for it to take place from the early stages of technology development and throughout the development process in order that advanced reactors are developed with the preferences, values, and needs of potential host communities in mind.

**Nuclear Waste Management**

There has been a recent and growing emphasis on consent-based siting of nuclear facilities broadly and nuclear waste management facilities specifically. This emphasis on consent, especially in the US, is borrowed from the successful pursuit of consent-based processes in Finland and Sweden in particular [37]. Yet even these forms of consent are based on the traditional TRL paradigm that calls for the development of a technology which is then followed by a search for users or 'hosts'. A well-known rhetorical argument made about nuclear waste management is that it is a technically solved problem. While this may in fact be true, the social elements of the problem remain unresolved. Even consent-based approaches, though a significant move in the right direction, may ultimately be insufficient because they advocate for presenting interested communities with a binary menu – the acceptance of a given, as-designed facility and waste disposal technology option, or its rejection, thus excluding the possibility of community input and engagement in the process of selecting a technology option and designing the waste management facility. The SRL framework calls for the latter approach and is premised on the expectation that engaging communities in an early and continuing conversation about how to develop a waste management facility is likely to lead to desirable outcomes for both communities and practitioners-policymakers – in the form of a potentially larger pool of interested communities and ultimately, the design and use of waste management facilities – for the short and long term – that are responsive to community preferences. As a move in this direction, the DOE, through its NEUP program in its most recent round of funding, awarded a large grant to an Integrated Research Project (which includes us, the authors of this paper). The IRP project is aimed at "Integrating socially led co-design into consent-based siting of interim storage facilities"[38].

## FUTURE WORK

The sociotechnical readiness level framework developed here calls for the direct participation of and engagement with user and community groups throughout the technology development process. As such these approaches are not currently used in technology development processes in the nuclear sector – or for that matter, for the development of complex sociotechnical systems. In future work we will seek to develop these approaches – by drawing on participatory and inclusive design processes that have been developed in the applied computer sciences (human-robot interaction and human-computer interaction), the applied design disciplines (urban and regional planning, architecture, and landscape architecture), and design research.

Additionally, in future work, we will develop supporting guidelines, processes, and questions to inform each stage of sociotechnical development such that the descriptions and assessment approaches for each level of sociotechnical development match those that have been codified in existing guidelines such as the Technology Readiness Assessment and Technology Maturation Process guide.

## CONCLUSIONS

This paper develops the sociotechnical readiness level framework which is grounded in a critique of the traditional technology readiness framework and conceptually motivated by the design research and science and technology studies scholarship. The development of the SRL framework has been motivated by the urgent need to develop advanced nuclear technologies – for energy as well as waste management – using sociotechnical approaches. While this is the particular motivation for developing this framework, it may equally apply to other large infrastructure sectors that are undergoing a reimagining and transition in the context of national and global infrastructure renewal efforts which are interlinked with the clean energy transition. We invite research and practitioner colleagues in other disciplines to explore applications of the SRL framework to their respective domains of work.

**APPENDIX: EQUITY AND JUSTICE IN A NUCLEAR CONTEXT**

It is important to conceptualize what we mean by equity and justice in a nuclear context. The five forms of justice mentioned earlier can be described as follows : (1) distributive justice concerns a fair distribution of benefits and burdens of a technology, (2) procedural justice concerns inclusivity and fairness of decision-making procedures, (3) recognition Justice concerns acknowledgment of past harms as well as inequalities, (4) restorative justice concerns using policy interventions to prevent or repair distributional, procedural, and recognition injustices and finally, (5) epistemic justice concerns the inclusion of diverse ways of knowing in the systems of knowledge and practice that inform how we conceptualize justice. Epistemic justice is especially important in the context of native and indigenous communities around the world who have traditional ecological knowledge which take many forms [39]. These forms of knowledge have historically been regarded as invalid or inferior to Western thought and knowledge. And yet, it is these indigenous and native forms of knowledge that were mobilized to tend to the environment and steward our ecosystems for centuries and far exceed many Western systems of thought in their longevity and temporal validity.

The different forms of justice described above – in particular, the first four – are frequently referred to in definitions of climate, energy, and environmental justice. It is also important to note that the descriptions of the different forms of justice above should be treated as working definitions, subject to reformulation in a place-based and context-specific manner. This is because equity and justice – climate, energy, and environmental – or even nuclear justice (if we treat it as a category of its own) – are likely to mean different things in different communities and their conceptualizations may vary within any single community. It is important to acknowledge and seek to understand this plurality where it exists.

**ACKNOWLEDGEMENTS**

An earlier version of this paper was submitted to the IHLRWM 2022 conference. We are grateful to the support from the CEM RISE3 initiative that supported the continuation of this work.

We are also grateful to colleagues at Deep Isolation for conversations that stimulated the development of the Sociotechnical Readiness Level framework.